\documentclass[twocolumn,secnumarabic,amssymb, nobibnotes, aps, prd]{revtex4-1}
\usepackage{graphics}
\usepackage{graphicx}
\usepackage{amsmath}

\begin{document}

\title{Classification of stable Dirac and Weyl semimetals with reflection and rotational symmetry}%

\author{$\mathrm{Zihao}\quad \mathrm{Gao}^{1,*},\quad \mathrm{Meng}\quad \mathrm{Hua}^{1,*},\quad \mathrm{Haijun}\quad \mathrm{Zhang}^{2},\quad \mathrm{Xiao} \quad \mathrm{Zhang}^{1,\dagger}$}%
\affiliation{1. Department of Physics, Sun-Yat-Sen University, Guangzhou, China}
\affiliation{2. National Laboratory of Solid State Microstructures, School of Physics, Collaborative Innovation Center of Advanced Microstructures, Nanjing University, Nanjing 210093, China.
$^*$These authors contributed equally to this work. $^{\dagger}$Correspondence and requests for materials should be addressed to Xiao Zhang (email: yngweiz@gmail.com).}
\begin{abstract}

Three dimensional (3D) Dirac semimetal is a novel state of quantum matter, characterized by the gapless bulk four-fold degeneracy near Fermi energy. Soon after its discovery, the classification of stable 3D Dirac semimetals with inversion and rotational symmetry have been studied. However, only ten out of thirty-two point groups have both inversion and rotational symmetry, and we need a more complete classification of stable 3D Dirac semimetals. Here we classify stable 3D Dirac semimetals with reflection symmetry and rotational symmetry in the presence of time reversal symmetry, which belong to seventeen different point groups. These systems include the systems preserving inversion symmetry except $\mathrm{C_{3i}}$. They have two classes of reflection symmetry, with the mirror plane parallel to rotation axis and the mirror plane perpendicular to rotation axis.  In both cases two types of Dirac semimetals are determined by four different reflection symmetries. The first type of Dirac semimetals will appear through accidental band crossing (ABC). The second type of Dirac semimetals have a Dirac point at a time reversal invariant momentum (TBC). We show that in both mirror parallel and perpendicular cases, $C_{2,3}$ symmetry can only protect stable Dirac points via TBC, while $C_{4,6}$ symmetry can have stable Dirac points as ABC or TBC. We further discuss that Weyl line nodes and Dirac semimetal can exist in Brillouin zone at the same time using $\mathrm{C_{4v}}$ symmetry as an example. Finally we classify Dirac line nodes and Weyl line nodes to show in which types of mirror plane they can exist.

\end{abstract}

\date{\today}%
\maketitle

\section{Introduction}
Dirac semimetals are new states of quantum matter. They have gap closing(Dirac points or Dirac line nodes) of conduction band and valence band which show pseudorelativistic physics of 3-dimensional(3D) Dirac fermions near the Fermi energy. Before the discovery of Dirac semimetals, the topological quantum states, such as topological insulator\cite{phystoday,zsc,kane,model} and graphene\cite{graphene1,graphene2}, can have 2-dimensional(2D) Dirac fermions. Different from topological insulators and superconductors, Dirac semimetals hold nontrivial features in the bulk states\cite{3d,2d,wz,dln1,book,chiral,z2,ln}. When the systems reach a quantum critical point between normal insulator and topological insulator, the accidental crossing of inverted bands will generate Dirac points in bulk\cite{hgte}.

To realize Dirac semimetal we can tune the chemical composition to the critical point of quantum phase transition\cite{hgte}, however those Dirac points are not stable. Soon stable 3D Dirac semimetals have been theoretically predicted\cite{cd3as2,wz} and observed experimentally in $\mathrm{Cd_3As_2}$ and $\mathrm{Na_3Bi}$\cite{ex1,ex2,ex3,ex4,ex5} by Angle-resolved photoemission spectroscopy (ARPES). In these materials there are two stable Dirac points in $k_z$ axis stabilized by rotational symmetry. While in $\beta$-cristobalite structure such like $\mathrm{BiO_2}$\cite{3d}, the Dirac points exist at a time reversal invariant momentum(TRIM). The unique electronic band structures of Dirac semimetals make them have unusual high mobility, oscillating quantum spin hall effect(QSHE) and giant diamagnetism \cite{ex3}. Recently, a new type of Dirac semimetal, with Dirac line nodes(DLN), has been proposed in $\mathrm{Cu_3NPd}$\cite{dln1,ln} and $\mathrm{LaN}$\cite{lan}.  DLN can exist in the system with or without spin-orbit coupling(SOC)\cite{fuliang}. At the same time, theoretical prediction shows that time reversal symmetry (TRS) breaking systems including $\mathrm{HgCr_2Se_4}$\cite{hgcr2se4} has Weyl nodes and a Weyl line node(WLN) in its mirror plane. Also systems with TRS breaking such as pyrochlore iridates\cite{wanxiangang} and $\mathrm{(CdO)_2(EuO)_2}$\cite{wln2} or  with TRS such as TaAs, NbAs, NbP and TaP\cite{daixiprx,hasan,hasan2,taas3,taas4,nbas,taas2,binghai,binghai2,ts1,ts2} have Weyl nodes.

Inspired by these work, we ask the question that which point group can protect Dirac semimetals. Yang and Nagaosa have classified Dirac points in physical systems preserving inversion symmetry\cite{yn}. However,
only ten out of thirty-two point groups have both inversion and rotational symmetry, and we need
a more complete classification of stable 3D Dirac points. Meanwhile, considering the critical role that reflection symmetry plays in newly predicted DLN and WLN semimetals, a classification of them by reflection symmetry becomes necessary as well. Here we classify 3D stable Dirac points in the systems preserving TRS, reflection symmetry and uniaxial rotational symmetry. Known that the reflection symmetry plays an important role in classification of topological phases\cite{cl1,cl2}, we first study the classification of reflection symmetries in space groups. Apart from point group $C_{3i}$, all the other point groups preserving inversion symmetry are covered in our discussion. Then we show that four different reflection symmetries can protect two different types of Dirac semimetals, Dirac semimetals created by ABC or by TBC. We further discuss the coexistence of Dirac points and Weyl line nodes through $k\cdot p$ theory using $\mathrm{C_{4v}}$ symmetry as an example. Finally we classify DLN and WLN to show in which types of mirror plane they can exist.

\begin{figure*}
\centering
 \includegraphics[width=0.7\textwidth]{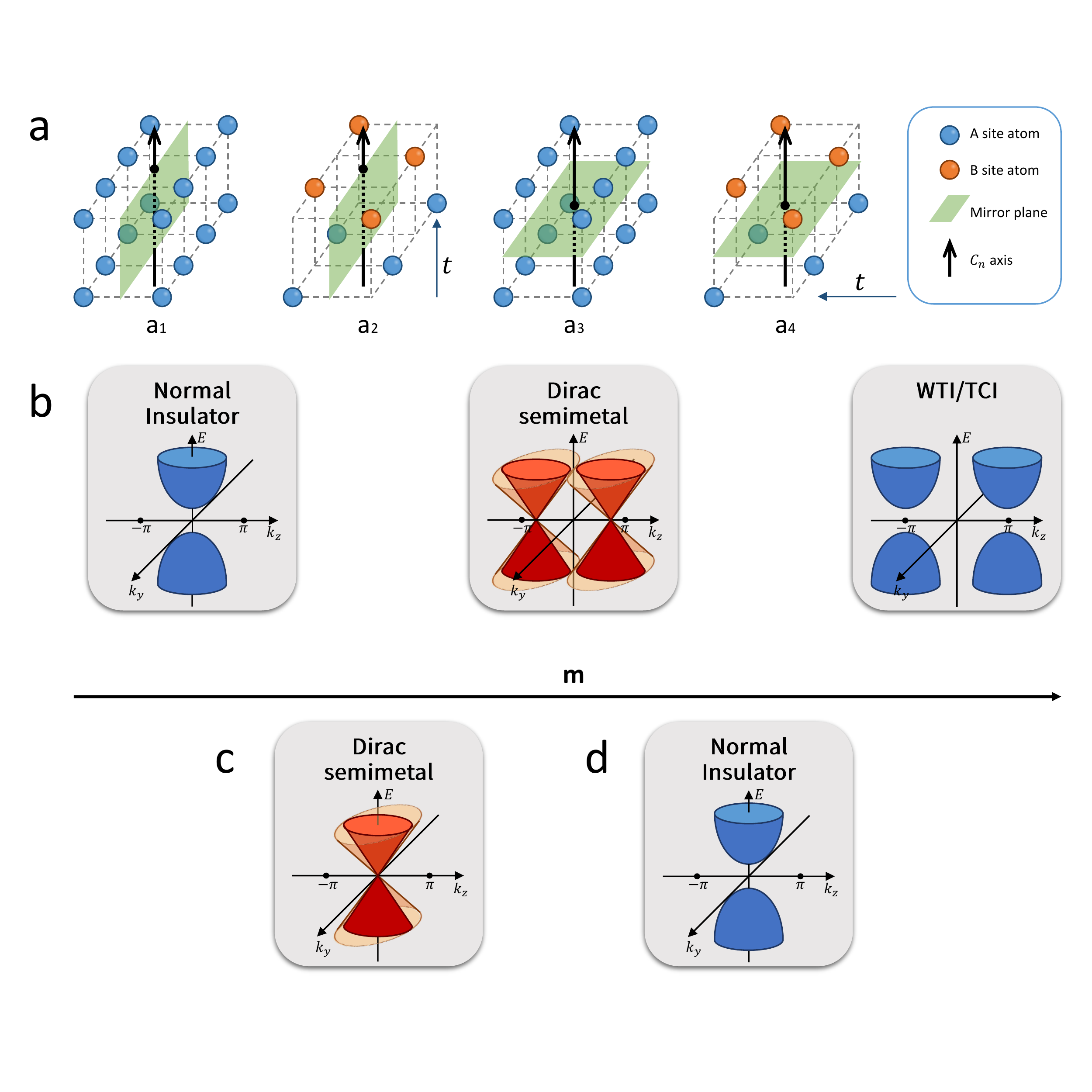}
\caption{($\mathbf{a}$)There are four kinds of reflection symmetries in space groups: the mirror plane parallel to rotation axis as shown in $\mathbf{a_1}$ and $\mathbf{a_2}$ and the mirror plane perpendicular to rotation axis as shown in $\mathbf{a_3}$ and $\mathbf{a_4}$. In both cases, for normal mirror plane $\mathbf{a_1}$ and $\mathbf{a_3}$, there is only one equivalent site. As for glide mirror plane $\mathbf{a_2}$ and $\mathbf{a_4}$, which is a reflection symmetry with a $t$ translation, there are two inequivalent sites in the lattice. (b),(d) correspond to the Dirac points of type ABC and TBC in TABLE \ref{tab:m} and TABLE \ref{tab:m2} respectively. (b)The phase transition determined by a control parameter $m$ and Dirac semimetal created by ABC. When $m$ is in a proper range, two Dirac points show up on $k_z$ axis. This phase lays between two gapped phase such as normal insulator and weak topological insulator (WTI) and topological crystalline insulator (TCI). (c)The Dirac semimetal phase is protected by crystalline symmetry. In this case, Dirac point can be found at TRIM. (d)When the conduction band and valence band have the same rotation eigenvalue, i.e. $p=q$, they will never cross each other because of strong level repulsion. 
}
\label{mirror}
\end{figure*}

\section{Results}
 Consider four energy bands which could be generated in a system preserving time reversal symmetry(TRS) and uniaxis rotational symmetry, we can describe them through a $4\times 4$ Hamiltonian in a very general form,
$$H=\left(
\begin{array}{cc}
 h_{\uparrow\uparrow}(\vec k) &  h_{\uparrow\downarrow}(\vec k) \\
  h_{\downarrow\uparrow}(\vec k) &  h_{\downarrow\downarrow}(\vec k) \\
\end{array}
\right)=\sum^3_{i,j=0}{a_{ij}(\vec k)\tau_i\sigma_j}$$
where $\sigma_i$ represents the spin space and $\tau_i$ represents orbital space. $h_{\sigma\sigma'}(\sigma=\uparrow,\downarrow)$ is a $2\times 2$ matrix and ${\uparrow\downarrow}$ represent opposite spin in $k_z$ direction. All the $a_{ij}(\vec k)$ are real functions and we can determine the parity of each coefficient $a_{ij}(\vec k)$ through TRS $ H(-\vec k)=T H(\vec k)T^{-1}$, where $T=i\sigma_y K$ . 
The system is invariant under $C_n$ rotational symmetry which gives $C_nH(\vec{k})C_n^{-1}=H(R_n\vec{k})$, where $R_n$ is a rotation operator for 3D n-fold rotation in k-space and the basis are chosen to be eigenstates of $C_n$. We choose the rotation axis as $k_z$ axis, therefore $R_nk_z=k_z$. Rotational symmetry suggests commutation relation $C_nH(k_z)C_n^{-1}=H(k_z)$ in $k_z$  between the Hamiltonian $H(k_z)$ and $C_n$ operators. Therefore we choose a set of bases to make $C_n$ a diagonal form $C_n=diag(\alpha_p,\alpha_q,\alpha_r,\alpha_s)$, where $\alpha_p=\exp[i\frac{2\pi}{n}(p+\frac{1}{2})]$, $p$ is the angular momentum for point group rotational symmetry and the effective angular momentum for a screw rotation. Here each basis is a $C_n$ rotation eigenstate and has a definite rotation eigenvalue $p+\frac{1}{2}$. In the presence of TRS, $C_n$ has only two independent basis whose rotation eigenvalues are $\alpha_p$ and $\alpha_q$ (see Appendix).

\begin{table*}
\centering
\caption{Classification table when $k_z$ is parallel to mirror plane. Dirac semimetals can be obtained in the systems with reflection and $C_n$ rotational symmetry. Here we choose $k_z$ as the rotation axis and assume that $yz$-plane is a mirror plane. $(p,q,r,s)$ can be regarded as a set of orbital angular momentum in $z$ direction and $j$ is the total angular momentum. For example, for $(p,q,r,s) = (3,2,0,1)$ in $C_4$ system, the $C_4$ rotation eigenvalues are $(e^{i\frac{7\pi}{4}},e^{i\frac{5\pi}{4}},e^{i\frac{1\pi}{4}},e^{i\frac{3\pi}{4}})$, which equal $(e^{-i\frac{1}{2}\frac{\pi}{2}},e^{-i\frac{3}{2}\frac{\pi}{2}},e^{i\frac{1}{2}\frac{\pi}{2}},e^{i\frac{3}{2}\frac{\pi}{2}})$. For compact presentation, we assume $n/2\leq q\leq p<n$ and consider the equivalence between $\{p,r\}$ and $\{q,s\}$. The leading order of $f_\pm,g_\pm,g_0+g_z,g_0-g_z$ are shown in the table. Each term should by multiplied by an coefficient function of $k_z$ respecting to the parity of $a_{ij}(k)$ when constructing the elements of the Hamiltonian. The bulk Dirac points are obtained through ABC when $M_x=\pm\tau_0/\tau_z\otimes\sigma_x$ and they are obtained through TBC when $M_x=\pm\tau_x/i\tau_y\otimes\sigma_x$. The material $\mathrm{Cd_3As_2}$ belongs to space group $I4_1cd$\cite{ex1}. }\label{tab:m}          
\begin{tabular}{l|l|l|l|llll|l|l|l}
        \hline
        $M_x$                & $C_n$     & (p,q,r,s) &Total $j$     & $f_\pm$ & $g_\pm$ & $g_0+g_z$ & $g_0-g_z$ & Dirac & \bf{Materials} & Dispersion in  \\ 
							 &  		 &         	 &		&        &		    &  		    &           & Type  &                &    $k_y$ direction  \\ \hline
        $\pm\tau_0/\tau_z$   & $C_2$ & -         &-& -       & -       & -         & -         & -          & -              & -                             \\ 
  & $C_3$ & -         &-& -       & -       & -         & -         & -          & -              & -                             \\ 
        ~                    & $C_4$     & (3,2,0,1) &$(\pm\frac{1}{2},\pm\frac{3}{2})$&$k_+$   &$k_\pm^2$   &$k_-$   &$k_+$      & ABC      & $\mathrm{Cd_3As_2}(I4_1cd)$ & Linear Dirac    \\ 
        ~                    & $C_6$     & (5,4,0,1) &$(\pm\frac{1}{2},\pm\frac{3}{2})$&$k_+$   &$k_-^2$   &$k_-$   &$k_\pm^3$    & ABC      & ~             & Linear Dirac                  \\ 
        ~                    & $C_6$     & (5,3,0,2) &$(\pm\frac{1}{2},\pm\frac{5}{2})$&$k_+^2$   &$k_\pm^3$   &$k_-$   &$k_+$    & ABC      & ~             & Linear Dirac                  \\ 
        ~                    & $C_6$     & (4,3,1,2) &$(\pm\frac{3}{2},\pm\frac{5}{2})$&$k_+$   &$k_+^2$   &$k_\pm^3$   &$k_+$    & ABC      & ~             & Linear Dirac                  \\ 
        \hline
		$\pm\tau_x$ & $C_2$     & -         &-& -       & -       & -         & -         & -          & -              & -                             \\ 
        ~                    & $C_3$     & (2,1,0,1) &$(\pm\frac{1}{2},\pm\frac{3}{2})$&$k_+$   & $k_+$  &$k_-$   & $k_+k_-$      & TBC      & ~             & Linear Dirac                  \\ 
        ~                    & $C_4$     & (3,2,0,1) &$(\pm\frac{1}{2},\pm\frac{3}{2})$&$k_+$   &$k_\pm^2$   &$k_-$   &$k_+$      & TBC      &               & Linear Dirac                  \\ 
        ~                    & $C_6$     & (5,4,0,1) &$(\pm\frac{1}{2},\pm\frac{3}{2})$&$k_+$   &$k_-^2$   &$k_-$   &$k_\pm^3$    & TBC      & ~             & Linear Dirac                  \\ 
        ~                    & $C_6$     & (5,3,0,2) &$(\pm\frac{1}{2},\pm\frac{5}{2})$&$k_+^2$   &$k_\pm^3$   &$k_-$   &$k_+$    & TBC      & ~             & Linear Dirac                  \\ 
        ~                    & $C_6$     & (4,3,1,2) &$(\pm\frac{3}{2},\pm\frac{5}{2})$&$k_+$   &$k_+^2$   &$k_\pm^3$   &$k_+$    & TBC      & ~             & Linear Dirac                  \\ 
		\hline
		$\pm i\tau_y$ 		 & $C_2$     & (1,1,0,0) &$(\pm\frac{1}{2},\pm\frac{1}{2})$&$k_+k_-$& $k_+$  & $k_\pm$& $k_\pm$       & TBC      & ~             & Linear Dirac                  \\ 
		~					 & $C_3$     & (2,2,0,0) &$(\pm\frac{1}{2},\pm\frac{1}{2})$&$k_+k_-$& $k_-$  & $k_-$  & $k_-$     	  & TBC      & ~             & Linear Dirac                  \\ 
        ~                    & $C_3$     & (2,1,0,1) &$(\pm\frac{1}{2},\pm\frac{3}{2})$&$k_+$   & $k_+$  & $k_-$  & $k_+k_-$      & TBC      & ~             & Linear Dirac                  \\ 
		~					 & $C_4$     & (3,3,0,0) &$(\pm\frac{1}{2},\pm\frac{1}{2})$&$k_+k_-$& $k_-$  & $k_-$  & $k_-$     	  & TBC      & ~             & Linear Dirac                  \\ 
        ~                    & $C_4$     & (3,2,0,1) &$(\pm\frac{1}{2},\pm\frac{3}{2})$&$k_+$   &$k_\pm^2$   &$k_-$   &$k_+$      & TBC      & ~             & Linear Dirac                  \\ 
		~					 & $C_4$     & (2,2,1,1) &$(\pm\frac{3}{2},\pm\frac{3}{2})$&$k_+k_-$& $k_+$  & $k_+$  & $k_+$     	  & TBC      & ~             & Linear Dirac                  \\ 
		~					 & $C_6$     & (5,5,0,0) &$(\pm\frac{1}{2},\pm\frac{1}{2})$&$k_+k_-$& $k_-$  & $k_-$  & $k_-$     	  & TBC      & ~             & Linear Dirac                  \\ 
        ~                    & $C_6$     & (5,4,0,1) &$(\pm\frac{1}{2},\pm\frac{3}{2})$&$k_+$   &$k_-^2$   &$k_-$   &$k_\pm^3$    & TBC      & ~             & Linear Dirac                  \\ 
        ~                    & $C_6$     & (5,3,0,2) &$(\pm\frac{1}{2},\pm\frac{5}{2})$&$k_+^2$ &$k_\pm^3$   &$k_-$   &$k_+$      & TBC      & ~             & Linear Dirac                  \\ 
		~					 & $C_6$     & (4,4,1,1) &$(\pm\frac{3}{2},\pm\frac{3}{2})$&$k_+k_-$&$k_\pm^3$&$k_\pm^3$  & $k_\pm^3$ & TBC      & ~             & Quadratic Dirac                   \\ 
        ~                    & $C_6$     & (4,3,1,2) &$(\pm\frac{3}{2},\pm\frac{5}{2})$&$k_+$   &$k_+^2$   &$k_\pm^3$   &$k_+$    & TBC      & ~             & Linear Dirac                  \\ 
		~					 & $C_6$     & (3,3,2,2) &$(\pm\frac{5}{2},\pm\frac{5}{2})$&$k_+k_-$& $k_+$  & $k_+$  & $k_+$     	  & TBC      & ~             & Linear Dirac                  \\ 
        \hline
    \end{tabular}
\\
\end{table*}

\begin{table*}
\caption{Classification table when $k_z$ is perpendicular to mirror plane. Dirac semimetals can be obtained in the systems with $C_n$ rotational symmetry and reflection symmetry. Here we choose $k_z$ as the rotation axis and assume that $xy$-plane is a mirror plane. $(p,q,r,s)$ can be regarded as a set of orbital angular momentum in z. For example, for $(p,q,r,s) = (2,0,1,3)$ in $C_4$ system, the $C_4$ rotation eigenvalues are $(e^{i\frac{5\pi}{4}},e^{i\frac{1\pi}{4}},e^{i\frac{3\pi}{4}},e^{i\frac{7\pi}{4}})$, which equal $(e^{-i\frac{3}{2}\frac{\pi}{2}},e^{i\frac{1}{2}\frac{\pi}{2}},e^{i\frac{3}{2}\frac{\pi}{2}},e^{-i\frac{1}{2}\frac{\pi}{2}})$. $j$ is the total angular momentum. For compact presentation, we assume $q\leq p<n$ and consider the equivalence between $\{p,r\}$ and $\{q,s\}$. The leading order of $f_\pm,g_\pm,g_0+g_z,g_0-g_z$ are shown in the table. Each term should be multiplied by an coefficient function of $k_z$ respecting to the parity of $a_{ij}(k)$ when constructing the elements of the Hamiltonian. The bulk Dirac points are obtained through ABC when $M_z=\pm\tau_0/\tau_z\otimes\sigma_z$ and they are obtained through TBC when $M_z=\pm\tau_x/i\tau_y\otimes\sigma_z$. The material $\mathrm{Cd_3As_2}$ belongs to space group $I4_1acd$.}\label{tab:m2}
\centering
         \begin{tabular}{lll|l|l|l|lll|l|l|l}
        \hline
        $M_z$           & $C_2$    & $P$             & $C_n$     & (p,q,r,s) &Total $j$  & $f_\pm$   & $g_\pm$   & $g_0\pm g_z$  & Dirac & Materials & Dispersion  \\
        ~               & ~        & ~               & ~	     & ~ 		 &~  		 & ~  		 & ~  		 & ~ 		     & type  & ~         & in $k_y$ \\ \hline
        $\tau_0/\tau_z$ & $\tau_0$ & $\tau_0/\tau_z$ & $C_2$     & -         & - 		 & -         & -         & -             & -          & -         & -          \\
        ~               & ~        & ~               & $C_4$     & (2,0,1,3) &$(\pm\frac{3}{2},\pm\frac{1}{2})$& $k^2_\pm$ & $k_+$     & 0         & ABC        & $\mathrm{Cd_3As_2}(I4_1acd)$         & Linear     \\
        ~               & ~        & ~               & $C_6$     & (2,0,3,5) &$(\pm\frac{5}{2},\pm\frac{1}{2})$& $k^2_+$   & $k^3_\pm$ & 0         & ABC        & ~         & Quadratic  \\
        ~               & ~        & ~               & $C_6$     & (3,1,2,4) &$(\pm\frac{5}{2},\pm\frac{3}{2})$& $k_+^2$   & $k_-$     & 0         & ABC        & ~         & Linear     \\
        ~               & ~        & ~               & $C_6$     & (4,0,1,5) &$(\pm\frac{3}{2},\pm\frac{1}{2})$& $k_-^2$   & $k_-$     & 0         & ABC        & ~         & Linear     \\
        ~               & $\tau_z$ & $\tau_z/\tau_0$ & $C_2$     & -         & - & -         & -         & -         & -         & -                 & -          \\
        ~               & ~        & ~               & $C_4$     & (1,0,2,3) &$(\pm\frac{3}{2},\pm\frac{1}{2})$& $k_+$     & $k^2_\pm$ & 0         & ABC        & ~         & Linear     \\
        ~               & ~        & ~               & $C_6$     & (1,0,4,5) &$(\pm\frac{3}{2},\pm\frac{1}{2})$& $k_+$     & $k_+^2$   & 0         & ABC        & ~         & Linear     \\
        ~               & ~        & ~               & $C_6$     & (2,1,3,4) &$(\pm\frac{5}{2},\pm\frac{3}{2})$& $k_+$     & $k_-^2$   & 0         & ABC        & ~         & Linear     \\
        ~               & ~        & ~               & $C_6$     & (3,0,2,5) &$(\pm\frac{5}{2},\pm\frac{1}{2})$& $k^3_\pm$ & $k^2_-$   & 0         & ABC        & ~         & Quadratic  \\
        \hline
        $i\tau_y$       & $\tau_0$ & $i\tau_y$       & $C_2/C_6$     & -         & - & -         & -         & -         & -                 & -         & -          \\
        ~               & ~        & ~               & $C_4$     & (2,0,1,3) &$(\pm\frac{3}{2},\pm\frac{1}{2})$& $k_\pm^2$ & $k_-$     & $k_+$         & TBC       & ~         & Linear     \\
        ~               & $\tau_z$ & $\tau_x$        & $C_2$     & (1,0,0,1) &$(\pm\frac{1}{2},\pm\frac{1}{2})$& $k_\pm$   & 0         & $k_\pm$     & TBC       & Distorted spinels\cite{dist}    & Linear     \\
        ~               & ~        & ~               & $C_4$     & -         & - & -         & -         & -         & -                & -         & -          \\
        ~               & ~        & ~               & $C_6$     & (3,0,2,5) &$(\pm\frac{5}{2},\pm\frac{1}{2})$& $k^3_\pm$ & 0         & $k_+$         & TBC       & ~         & Linear     \\
        ~               & ~        & ~               & $C_6$     & (4,1,1,4) &$(\pm\frac{3}{2},\pm\frac{3}{2})$& $k^3_\pm$ & 0         & $k^3_\pm$  & TBC       & ~         & Cubic      \\
        \hline
        $\tau_x$        & $\tau_0$ & $\tau_x$        & $C_2/C_6$ 	 & -         & - & -         & -         & -         & -               & -         & -          \\
        ~               & ~        & ~               & $C_4$     & (2,0,1,3) &$(\pm\frac{3}{2},\pm\frac{1}{2})$& $k_\pm^2$ & 0         & $k_+$         & TBC       & $\mathrm{BiO_2}$         & Linear     \\
        ~               & $\tau_z$ & $i\tau_y$       & $C_2/C_4$ 	 & -         & - & -         & -         & -         & -         & -                 & -          \\
        ~               & ~        & ~               & $C_6$     & (3,0,2,5) &$(\pm\frac{5}{2},\pm\frac{1}{2})$& $k_\pm^3$ & $k_-^2$   & $k_+$         & TBC       & ~         & Linear     \\
        \hline
    \end{tabular}
\\
\end{table*}

After applying TRS and rotational symmetry, the Hamitonian in $k_z$ axis becomes a diagonal form with the above basis  
\begin{equation}\label{eq1}
H(k_z)=a_{00}+a_{03}\sigma_3+a_{33}\tau_3\sigma_3+a_{30}\tau_3
\end{equation}
The Dirac points are created only when $a_{03,33,30}(k_z,m)=0$, where $m$ is a control parameter. There are three equations and two variables, so we need additional symmetry constraints to guarantee these equations having at least one solution that can generate stable Dirac points. We impose reflection symmetry into the systems with uniaxial rotational symmetry and TRS, and show the conditions where Dirac points exist.

\begin{table*}
\caption{Classification table of topological phase for $D_{3h}$. In the presence of $C_3$ symmetry, Dirac semimetals can only be obtained in the systems with $M_z=i\tau_y\otimes\sigma_z$. $H(k_z)$ is the Hamiltonian on $k_z$ axis after considering reflection symmetry but without constraints by rotation symmetry. }\label{tab:d3h}
\begin{tabular}{l|l|l|l|l|l}
	\hline
	Mirror  & $H(k_z)$ & $C_n$ 	    & Possible 	  &Total $j$ & Dirac\\
	operator & 	    & Constraint & $(p,q,r,s)$ & & type\\ \hline
	$\tau_0$ & $h_{\uparrow\uparrow}=a_{10}\tau_x+a_{23}\tau_y+a_{30}\tau_z$ &  $p\neq q\neq r$ & -- & -- & --\\ 
			  & $h_{\uparrow\downarrow}=a_{01}-ia_{02}+(a_{11}-ia_{12})\tau_x+(a_{31}-ia_{32})\tau_z$ &  $q\neq s$ & & \\
	\hline
	$\tau_x$ & $h_{\uparrow\uparrow}=a_{10}\tau_x+a_{20}\tau_y+a_{33}\tau_z$ &  $p\neq q\neq r$ &-- &--& \\ 
			  & $h_{\uparrow\downarrow}=a_{01}-ia_{02}+(a_{11}-ia_{12})\tau_x-i(a_{21}-ia_{22})\tau_y$ & $q\neq s$ & & & \\
	\hline
	$i\tau_y$ & $h_{\uparrow\uparrow}=a_{13}\tau_x+a_{23}\tau_y+a_{33}\tau_z$ & $p\neq q$ & (2,0,0,2) &$(\pm\frac{1}{2},\pm\frac{1}{2})$ & TBC\\ 
			   & $h_{\uparrow\downarrow}=a_{01}-ia_{02}$ 						   & 	        & (2,1,0,1) &$(\pm\frac{1}{2},\pm\frac{3}{2})$ & TBC \\
	\hline
	$\tau_z$ &$h_{\uparrow\uparrow}=a_{13}\tau_x+a_{20}\tau_y+a_{30}\tau_z$  &  $p\neq q\neq r$ &-- & -- & \\ 
			  & $h_{\uparrow\downarrow}=a_{01}-ia_{02}-i(a_{21}-ia_{22})\tau_y+(a_{31}-ia_{32})\tau_z$ & $q\neq s$ & & & \\
	\hline
\end{tabular}
\end{table*}

\begin{figure*}[]
\centering
{\includegraphics[width=1\textwidth]{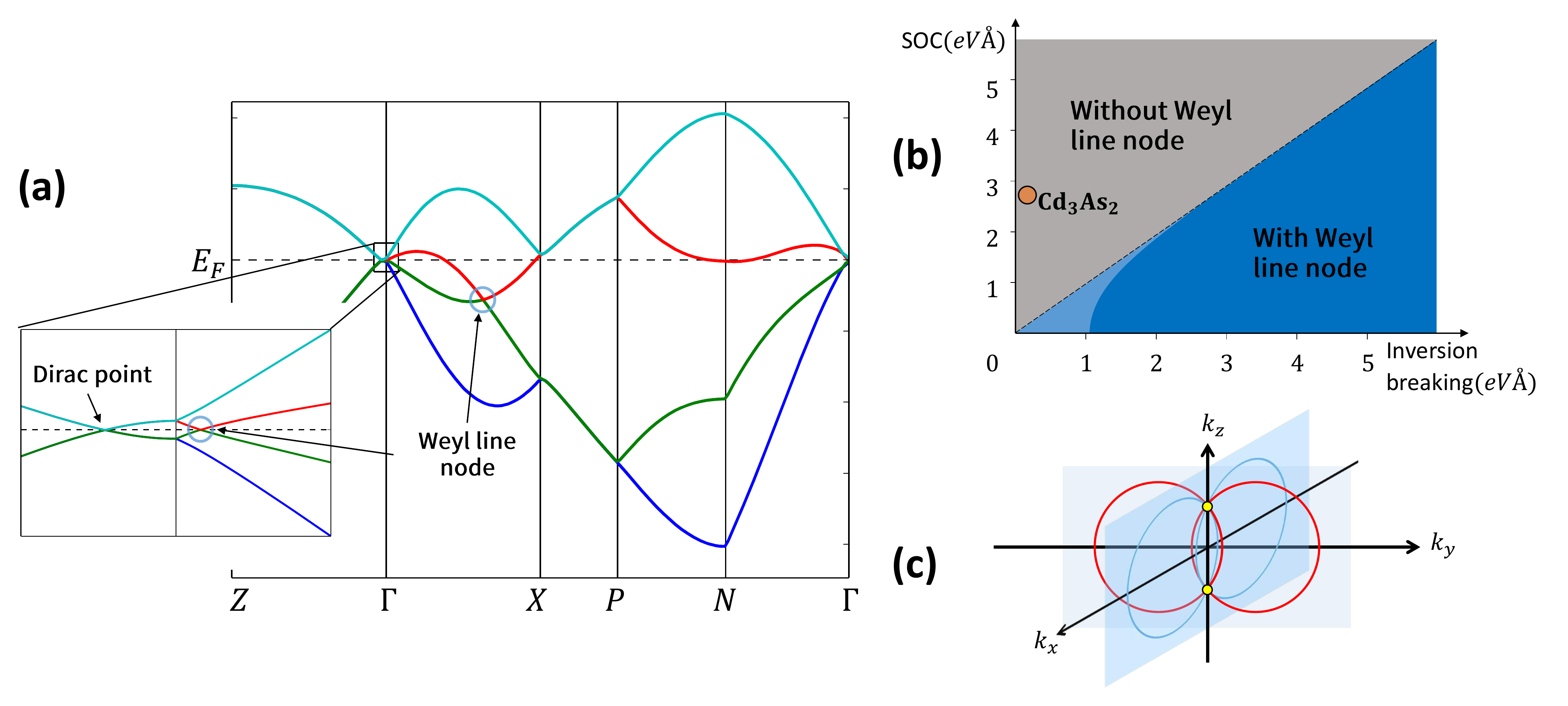}}
\caption{(a) The electronic structure of $C_{4v}$ system with SOC. The Dirac points and Weyl line nodes can exist simultaneously. The parameters are listed in the Appendix TABLE \ref{tpara}. (b)Phase diagram with respect to $B_0$, $D_0$ and $A_0$ of $C_{4v}$ system results from the Hamiltonian [Equation(\ref{He})] of $k\cdot p$ perturbation method. For all other parameters we fix them using the values fixed for $\mathrm{Cd_3As_2}$\cite{ex2}(see Appendix TABLE \ref{tpara}).  The inversion breaking term is $B_0$ and SOC is described by $D_0$. Weyl line nodes exist in the blue areas, among which the dark blue area represents the situation when the parameter $A_0=−0.06eV$($\mathrm{Cd_3As_2}$)\cite{ex2} and the light blue area (partially covered by the dark one) for $A_0=−0.00922eV$($\mathrm{HgTe}$)\cite{hgtepara}. (c)Schematic diagram of the distribution of Dirac points(yellow points) and Weyl line nodes(red or blue circles) in Brillouin zone. Dirac points  locate on the interception of four Weyl line nodes.}
\label{fig:phase}
\end{figure*}

\subsection{Classification of reflection symmetry}
Among the point groups, the reflection symmetries can be distinguished in two classes by the relative positions between mirror plane and rotation axis $k_z$.  
In the first class, $k_z$ axis parallels the mirror plane, and the system doesn't preserve inversion symmetry which correspond to point groups $C_{2v}, C_{3v}, C_{4v}, C_{6v}$,$D_{2d}$ and $T_d$. In these systems, reflection symmetry can be a point group symmetry as shown in FIG. \ref{mirror}.($\mathrm{a_1}$) or a nonsymmorphic glide plane symmetry which is the combination of  a reflection and a translation $t$ as shown in FIG. \ref{mirror}.($\mathrm{a_2}$).  Neupane et. al\cite{ex1} realized Dirac semimetal on $\mathrm{Cd_3As_2}$ which belongs to nonsymmorphic space group $I4_1cd$ ($C_{4v}$). Whereas in the second class, when $k_z$ axis is perpendicular to the mirror plane, inversion symmetry will emerge through the combination of reflection symmetry and $C_2,C_4$ or $C_6$ rotational symmetry\cite{yn,ti}. Here the point group reflection symmetry is shown in FIG. \ref{mirror}.($\mathrm{a_3}$) and the glide plane symmetry in nonsymmorphic space group is shown in FIG. \ref{mirror}.($\mathrm{a_4}$).

 Mirror operator is an inversion operation followed by a $C_2$ rotation whose rotation axis perpendicular to the mirror plane. It should satisfy the following constraints: (1)$[M,T]=0$, (2)$M^+M=1$, (3)$MM=e^{i\phi}$. If we set mirror plane as $yz$-plane or $xy$-plane, the reflection symmetry operator will have the form:
(A)$M_k=\pm\tau_0\otimes i\sigma_k$,  (B)$M_k=\pm\tau_z\otimes i\sigma_k$, (C)$M_k=\pm\tau_x\otimes i\sigma_k$and (D)$M_k=\pm i\tau_y\otimes i\sigma_k$($k=x,z$, see Appendix), among which (C, D) represent glide mirror symmetry.
For example, with the basis containing only equivalent sites[see FIG. \ref{mirror}.($\mathrm{a_1}$)] $|P_{A+\uparrow}\rangle, |P_{A-\uparrow}\rangle,|P_{A-\downarrow}\rangle, |P_{A+\downarrow}\rangle$, a reflection operation along $yz$-plane writes $M_x=-\tau_0\otimes i\sigma_x$. And if we set the basis as $|P_{A+\downarrow}\rangle, |P_{B+\downarrow}\rangle, -|P_{A-\uparrow}\rangle, -|P_{B-\uparrow}\rangle$, where $A$/$B$ are two inequivalent sites for the same kind of atom [FIG. \ref{mirror}.($\mathrm{a_2}$)], then the reflection operation along $yz$-plane with an additional interchanging of $A$ and $B$ sites writes $M_x=\tau_x\otimes i\sigma_x$(see Appendix). 
Note that the basis in our framework is the single atom basis after considering spin-orbit coupling(SOC). 

Owing to the TRS and rotational symmetry, the Hamiltonian in $k_z$ axis has been constrained as Eq. (\ref{eq1}). After we impose reflection symmetry, the $H(k_z)$ can be further constrained and will reveal Dirac semimetal phase. As shown in Appendix, different mirror operator will hold different type of Dirac semimetal. When the mirror operator is $M_k=\tau_0/\tau_z\otimes\sigma_k(k=x,z)$, only $a_{30}(k_z,m)$ survives. $a_{30}(k_z)$ is an even function respect to $k_z$, so $a_{30}(k_z,m)\approx M_0-M_1k_z^2$ for the leading order. Two Dirac points will emerge in the $k_z$ axis at $k_z=\pm\sqrt{M_0/M_1}$. This kind of Dirac semimetal is created through the two bands accidentally crossing each other when the conduction band and valence band have different rotation eigenvalues ($p\neq q$). It can be understood as a phase between normal insulator and weak topological insulator\cite{wti2,ti2}/topological crystalline insulator\cite{tci,tci2} [FIG. \ref{mirror}.(b)]. When mirror operator is $M_k=i\tau_y/\tau_x\otimes\sigma_k$, only $a_{33}(k_z,m)$ survives which is an odd function respect to $k_z$. The leading order is $a_{33}=M_2 k_z$. There is one Dirac point at TRIM generated by band crossing (TBC). Under this scenario, the Dirac points are created and stabilized by the crystalline symmetry [FIG. \ref{mirror}.(c)(d)]. 
We will use tables below to show the physical properties of all kinds of Dirac semimetals.

\subsection{Classification table}
If we impose reflection symmetry, TRS and rotational symmetry to the Hamiltonian in $k_z$ axis, we will have the criteria for basis's rotational eigenvalues to obtain Dirac semimetal phase. The classification of Dirac semimetals when the mirror plane is parallel to $k_z$ axis is shown in TABLE \ref{tab:m}. $C_2$ and $C_3$ rotational systems can only generate Dirac points via TBC. $C_4$ and $C_6$ symmetries can protect Dirac semimetal phase in the presence of all those four reflection symmetries. 

The classification of 3D Dirac semimetals when the mirror plane is perpendicular to $k_z$ axis is demonstrated in TABLE \ref{tab:m2}. Inversion symmetry can emerge through $P=C_2 m_z$, where $C_2$ is a two-fold rotation along $k_z$ axis in systems preserving $C_2, C_4, C_6$ symmetry. Yang and Nagaosa \cite{yn} have already considered unitary inversion operator $P=\tau_0,\tau_x,\tau_z$. We show that in TABLE \ref{tab:m2} the same results hold when inversion operators are unitary. However, the mirror operators can also generate the antiunitary inversion operator $P=i\tau_y$. $P=i\tau_y$ is an inversion operator with a translation in nonsymmorphic space group. It will produce a phase factor after being applied twice.
One special case is the $D_{3h}$ group, which does not involve inversion symmetry. As shown in TABLE \ref{tab:d3h}, only $M_z=i\tau_y\otimes\sigma_z$ can protect a Dirac point at TRIM with the combination of rotation eigenvalues $p=2,q=0$ or $p=2,q=1$.

\subsection{The coexistence of Dirac points and Weyl line nodes 
}
For now we can construct Dirac points in the systems preserving reflection symmetry but without inversion symmetry. It is known that  breaking inversion symmetry can produce Weyl points or Weyl line nodes. So we expect in some cases Dirac points and Weyl line nodes can exist simultaneously with SOC. We illustrate this distinctive properties in materials with $C_{4v}$ point group like $\mathrm{Cd_3As_2}$ through $k\cdot p$ perturbation method (see Appendix). Note that we choose mirror operator as $m_x=\tau_z\otimes i\sigma_x$ with p=3,q=2,r=0,s=1 corresponding to the 3rd row of TABLE \ref{tab:m}. The detailed calculations are shown in Appendix. By choosing some proper parameters, the band structure shown in FIG. \ref{fig:phase}.(a) exhibits two Dirac points along $\Gamma-Z$ direction and two Weyl line nodes in $yz$-plane and $xz$-plane. FIG. \ref{fig:phase}.(b) shows the phase transition between different topological phases. Obviously the phase transition of Weyl line nodes is independent of the emergence of Dirac points in $k_z$ direction[FIG. \ref{fig:phase}.(c)]. The simultaneous appearance of Dirac points and Weyl line nodes indicates a new class of topological phase with time reversal symmetry and SOC.

As illustrated in FIG. \ref{fig:phase}.(a) this $k\cdot p$ Hamiltonian can have different topological phases. These phases depend on the SOC term $D_0$ and inversion breaking term $B_0$. FIG. \ref{fig:phase}.(b) shows the phase diagram respecting to $B_0$ and $D_0$. There are two different phases when changing parameters. In the gray area, the systems break inversion symmetry but can not protect Weyl semimetal. In the dark blue area, the conduction band and valence band will cross to form a Weyl line node with $A_0=-0.06eV$. When $A_0=-0.00922eV$ the dark blue area extend to the light blue area . When the inversion breaking term $B_0\approx 0$, the system will not have Weyl line nodes phase on the $D_0$ axis, but it can protect Dirac semimetal phase just like in $\mathrm{Cd_3As_2}(I4_1acd)$. 
This phase diagram is independent of the creation of Dirac semimetal phase in bulk band structure. 

Numerical calculation shows that the two crossing bands of Weyl line nodes in blue areas have different mirror eigenvalues, which protects the gap closing on the Weyl line nodes.

Weyl line nodes are protected when the crossing bands have different mirror eigenvalues due to the absence of level repulsion. As is shown in FIG. \ref{linenode}.(c)(d), Weyl line nodes may emerge on various kinds of mirror plane.

\subsection{Dirac and Weyl line nodes in mirror plane}


Recently proposed by Fang, et al\cite{fuliang},
Dirac line nodes in mirror plane emerge in systems with inversion symmetry, time reversal symmetry and reflection symmetry. The combination of inversion operation and reflection operation generates $C_2$ rotation perpendicular to mirror plane, which is covered by our framework. Due to the combination of inversion symmetry and TRS, two bands related by TRS actually stick together and make up a two-fold degenerate band. 

\begin{figure}
\centering
\includegraphics[width=0.5\textwidth]{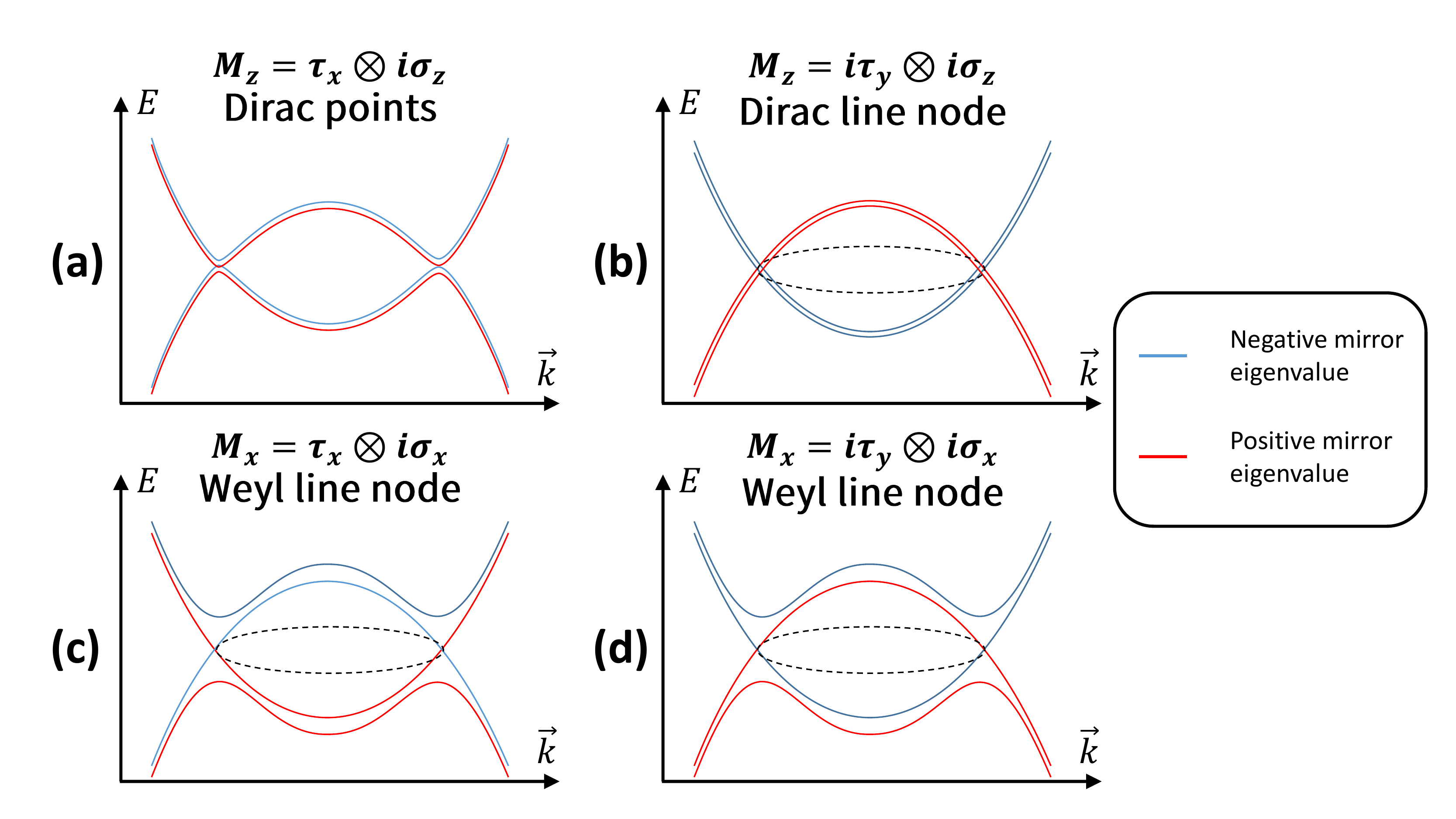}
\caption{Dirac or Weyl line nodes protected by reflection symmetry. Only $i\tau_y\otimes\sigma_z$ can protect Dirac line nodes while all kinds of mirror plane can protect Weyl line nodes. 
($\mathbf{a}$)($\mathbf{b}$)\cite{fuliang} are inversion preserving systems where Dirac line nodes may emerge and ($\mathbf{c}$)($\mathbf{d}$) are inversion breaking systems where Weyl line nodes may emerge. 
($\mathbf{a}$) For mirror operators other than $i\tau_y\otimes i\sigma_z$(e.g. $\tau_x\otimes i\sigma_z$), two bands with different mirror eigenvalues compose conduction (valence) band. Apart from two Dirac points, a gap between conduction and valence bands will open due to level repulsion. 
($\mathbf{b}$) For mirror $M_z=i\tau_y\otimes i\sigma_z$, conduction (valence) band consists of two bands with the same mirror eigenvalues. There isn't level repulsion between conduction and valence bands and the crossing of two bands generates a Dirac line node. The existence and classification of DLN by different reflection symmetries are shown in TABLE \ref{surdtable}.
($\mathbf{c}$)($\mathbf{d}$) For various kinds of mirrors (e.g. $i\tau_y\otimes i\sigma_x$ and $\tau_x\otimes i\sigma_x$), Weyl line nodes can be protected as long as the crossing bands have different mirror eigenvalues and so level repulsion doesn't happen (TABLE \ref{surdtable}).
}
\label{linenode}
\end{figure}

When we study the Hamiltonian in the mirror plane, the mirror symmetry give $MH(k_x,k_y)|_{k_z=0}M^{-1}=H(k_x,k_y)|_{k_z=0}$. Therefore, if the conduction (valence) band consists of two bands with different mirror eigenvalues, the bands with the same (positive or negative) mirror eigenvalue from the conduction and valence band will have strong level repulsion and open up a gap (FIG. \ref{linenode}.(a)(b)).
 Otherwise when TRS-partner bands have the same mirror eigenvalues, the conduction band will have a different mirror eigenvalue from the valence band, and the level repulsion will be relaxed. The band crossing between them will create Dirac line nodes in the mirror plane. Thus Dirac line nodes in mirror plane can only be protected by the nonsymmorphic $M_z=\pm i\tau_y\otimes i\sigma_z$ (FIG. \ref{linenode}.(a)(b) and TABLE \ref{surdtable}).

Different from Weyl line nodes in $\mathrm{HgCr_2Se_4}$\cite{hgcr2se4} and $\mathrm{(CdO)_2(EuO)_2}$\cite{wln2}, we construct Weyl line nodes in the presence of TRS. Weyl line nodes are protected when the crossing bands have different mirror eigenvalues. As is shown in FIG. \ref{linenode}.(c)(d) and TABLE \ref{surdtable}, Weyl line nodes may emerge on various kinds of mirror plane because of its two-fold degeneracy nature. Whereas Dirac line nodes in mirror plane can only be protected by $M_z=\pm i\tau_y\otimes i\sigma_z$ which satisfies the requirement of mirror eigenvalues  above. Note that DLN and WLN protected by mirror symmetry also apply to two-dimensional systems, because in three-dimensional systems we just take a plane into consideration by fixing $k_z$.

\section{Conclusion}

\begin{table} 
\centering
\caption{The possible protected semimetal phases by reflection symmetry. The mirror operators with $\sigma_x$ stands for mirror parallel to rotation axis and $\sigma_z$ for mirror perpendicular to rotation axis. Dirac line nodes in mirror plane is denoted by $\mathrm{DLN}$.}
\label{surdtable}
\begin{tabular}{c|cccc}
\hline
Mirror opertor 		     &  ABC 	& TBC 	 & WLN 	  & DLN  \\ \hline
$\tau_0\otimes i\sigma_x$ & $\surd$ &        & $\surd$ & 		\\
$\tau_x\otimes i\sigma_x$ &  	  & $\surd$ & $\surd$ & 		   \\
$i\tau_y\otimes i\sigma_x$ &  	  & $\surd$ & $\surd$ & 		   \\
$\tau_z\otimes i\sigma_x$ & $\surd$ & 	   & $\surd$ & 		   \\
$\tau_0\otimes i\sigma_z$ & $\surd$ &		   & $\surd$ & 		\\
$\tau_x\otimes i\sigma_z$ & 		  & $\surd$ & $\surd$ & 		   \\
$i\tau_y\otimes i\sigma_z$ & 	  & $\surd$ & $\surd$ & $\surd$ \\
$\tau_z\otimes i\sigma_z$ & $\surd$ &        & $\surd$ & 		  \\\hline
\end{tabular}
\end{table}

In this work we show that reflection symmetry can protect Dirac semimetal phase with or without inversion symmetry. We have classified Dirac semimetal in the systems preserving reflection symmetry, rotational symmetry and TRS. TABLE \ref{surdtable} can be referred to for an overall possible protection of semimetal phases by reflection symmetry. There are two kinds of Dirac semimetals created via ABC and TBC. 
In $C_2$ and $C_3$ rotation invariant systems, Dirac semimetals can be created only via TBC. Whereas in $C_4$ and $C_6$ systems, Dirac semimetals can be created via not only ABC but also TBC. 
We also found that in $C_{4v}$ point group system, Dirac semimetal phase can coexist with Weyl line nodes. Finally we show that DLN in mirror plane can be protected only by $M_z=i\tau_y\otimes i\sigma_z$ and WLN can be protected by any mirror operator. These new classes of Dirac semimetals in inversion breaking and preserving systems can guide the search for novel materials with exotic quantum properties\cite{dw}.

\section{Acknowledgments}
Xiao Zhang is support by the National Natural Science Foundation of China (No.11404413). Meng Hua and Zihao Gao acknowledge financial support from Yat-sen school, Sun Yat-sen University.

\section{Appendix}
\subsection{The constraint on Hamiltonian by TRS and rotational symmetry}
In this section we first constrain the Hamiltonian with TRS and rotational symmetry. 
 Then we impose rotational symmetry $C_n$ to the whole system to determine the leading order of the elements in $h_{\uparrow\uparrow}(\vec k)$ and $h_{\uparrow\downarrow}(\vec k)$. The Hamiltonian and the basis we use here are the same as those in the main text. Time reversal symmetry will give constraints to the Hamitonian $H(\vec k)$: $ H(-\vec k)=T H(\vec k)T^{-1}$, where the time reversal operator $T=i\sigma_y K$. We will have\cite{yn} 
$$ H=\left(
\begin{array}{cc}
 h_{\uparrow\uparrow}(\vec k) &  h_{\uparrow\downarrow}(\vec k) \\
 -h^*_{\uparrow\downarrow}(-\vec k) &  h^*_{\uparrow\uparrow}(-\vec k) \\
\end{array}
\right)$$

At the same time, we know the parity of each coefficient with respect to momentum $a_{01,02,03,11,12,13,20,31,32,33}(-\vec k)=-a_{01,02,03,11,12,13,20,31,32,33}(\vec k)$ and $a_{00,10,21,22,23,30}(-\vec k)=a_{00,10,21,22,23,30}(\vec k)$. 
We set the rotation axis as $k_z$ and choose the eigenstates of the rotation operator $C_n$ as the basis of matrices. Then the matrix representation of $C_n$ is

\begin{equation}
\begin{split} 
C_n
&=\mathrm{diag}[\alpha_p,\alpha_q,\alpha_r,\alpha_s]\\
&=\left(\begin{array}{cccc}
e^{i\frac{2\pi}{n}(p+1/2)}&0&0&0\\
0&e^{i\frac{2\pi}{n}(q+1/2)}&0&0\\
0&0&e^{i\frac{2\pi}{n}(r+1/2)}&0\\
0&0&0&e^{i\frac{2\pi}{n}(s+1/2)}\\
\end{array}\right)\\
&=\left(\begin{array}{cc}
e^{i\pi(\frac{1+p+q}{n}+\frac{p-q}{n}\tau_z)}&0\\
0&e^{i\pi(\frac{1+r+s}{n}+\frac{r-s}{n}\tau_z)}\\
\end{array}\right)
\end{split}
\end{equation}
where $p,q,r,s\in\{0,1,...,n-1\}$ and can be regarded as orbital angular momentum of different states. In general, $C_n$ commute with TRS $[C_n,T]=0$, thus $p$ and $r$, $q$ and $s$ are related by:
\begin{equation}
\begin{split}
\alpha_p=\bar{\alpha_r}&,\alpha_q=\bar{\alpha_s}\\
\exp[i\frac{2\pi}{n}(p+r+1)]=1&,\exp[i\frac{2\pi}{n}(q+s+1)]=1
\end{split}
\end{equation}

Next we derive the constraint relations between rotational symmetry and elements of the Hamiltonian. The $2\times 2$ block Hamiltonian $h_{\uparrow\uparrow},h_{\uparrow\downarrow}$ can be expanded in the following way\cite{multi}:
\begin{equation}
\begin{split}
h_{\uparrow\uparrow}(\vec k)&=f_0(\vec k)+f_+(\vec k)\tau_++f_+^*(\vec k)\tau_-+f_z(\vec k)\tau_z\\
h_{\uparrow\downarrow}(\vec k)&=g_0(\vec k)+g_+(\vec k)\tau_++g_-(\vec k)\tau_-+g_z(\vec k)\tau_z
\end{split}
\end{equation}
where $\tau_\pm=\tau_x\pm i\tau_y$, $f_{0,z}$ are real functions and $f_+,g_0,g_z,g_\pm$ are complex functions.
Then the rotational symmetry $C_nH(k_\pm,k_z)C_n^{-1}=H(k_\pm e^{\pm i2\pi/n},k_z)$ gives the constraints of elements of the Hamiltonian:
\begin{equation}\label{eqcr}
\begin{split}
f_z(k_\pm,k_z)&=f_z(k_\pm e^{\pm i2\pi/n},k_z)\\
\exp\left[i\frac{2\pi}{n}(p-q)\right]f_+(k_\pm,k_z)&=f_+(k_\pm e^{\pm i2\pi/n},k_z)\\
\exp\left[i\frac{2\pi}{n}(p-r)\right]g_{0+z}(k_\pm,k_z)&=g_{0+z}(k_\pm e^{\pm i2\pi/n},k_z)\\
\exp\left[i\frac{2\pi}{n}(q-s)\right]g_{0-z}(k_\pm,k_z)&=g_{0-z}(k_\pm e^{\pm i2\pi/n},k_z)\\
\exp\left[i\frac{2\pi}{n}(q-r)\right]g_\pm(k_\pm,k_z)&=g_\pm(k_\pm e^{\pm i2\pi/n},k_z)
\end{split}
\end{equation}
where $k_\pm=k_x\pm ik_y$, $g_{0\pm z}=g_0\pm g_z$.

On $k_z$ axis, these constraints become,
\begin{equation}\label{eqcrz}
\begin{split}
f_z(k_z)&=f_z(k_z)\\
\exp\left[i\frac{2\pi}{n}(p-q)\right]f_+(k_z)&=f_+(k_z)\\
\exp\left[i\frac{2\pi}{n}(p-r)\right]g_{0+z}(k_z)&=g_{0+z}(k_z)\\
\exp\left[i\frac{2\pi}{n}(q-s)\right]g_{0-z}(k_z)&=g_{0-z}(k_z)\\
\exp\left[i\frac{2\pi}{n}(q-r)\right]g_\pm(k_z)&=g_\pm(k_z)
\end{split}
\end{equation}
After considering all other symmetric relations (e.g. reflection symmetry in \ref{CHRS}), if non-diagonal $f,g$ terms aren't eliminated on $k_z$ axis, the corresponding (p,q,r,s) pairs ($f_+$ to $p,q$, $g_0+g_z$ to $p,r$, $g_0-g_z$ to $q,s$, $g_\pm$ to $q,r$) should be unequal to obtain Dirac points. For example, if $f_+$ exists on $k_z$ axis, we should apply $p\neq q$ or a gap will open on $k_z$ axis.

In order to get the dispersion relation near Dirac points, we should change the $f,g$ terms into a more explicit form. The matrix elements can be expanded as polynomial\cite{multi}:
\begin{equation}
f(k_+,k_-)=\sum_{n_1,n_2}{A_{n_1n_2}k_+^{n_1}k_-^{n_2}}.
\end{equation}
Combined with the constraint relations (\ref{eqcr}), we obtain
\begin{equation}
\begin{split}
e^{i2\pi(p-q)/n}f(k_+,k_-)=f(k_+ e^{i2\pi/n},k_- e^{-i2\pi/n})\\
=\sum_{n_1,n_2}{\exp\left[i\frac{2\pi}{n}(n_1-n_2)\right]A_{n_1n_2}k_+^{n_1}k_-^{n_2}}
\end{split}
\end{equation}
where $A_{n_1n_2}$ is an complex coefficient. To satisfy above equations, the phase factors must cancel each other, i.e. $n_1-n_2=(p-q)$ mod $n$. We choose the leading order terms to complete TABLE \ref{tab:m} and TABLE \ref{tab:m2}. For example, in $C_4$ system with $p=3$ and $q=2$, the constraint relation of $f_+$ term is,
\begin{equation}
\exp\left(i\frac{2\pi}{n}\right)f_+(k_\pm,k_z)=f_+(k_\pm e^{\pm i2\pi/n},k_z)
\end{equation}
Obviously the leading order term is given by $n_1=1,n_2=0$, and so we can replace the $f_+$ term by $A_{1,0}k_+$, neglecting the higher order terms.

\subsection{Details about the classification of reflection symmetry operators}

We study the general classification of reflection symmetry operators. The matrix representation of reflection symmetry can be decomposed to orbital space and angular momentum space. In spin space, mirror operator is a two-fold rotation perpendicular to the mirror plane $\hat n$: $R_{\pi/2}(\hat n)=e^{-i\hat{\sigma}j\cdot\hat{n}\pi/2}$, where $j$ indicates the half-integer spin momentum. At the same time, mirror operators should satisfy the following constraints: (1)$[M,T]=0$, (2)$M^+M=1$, (3)$MM=e^{i\phi}$. 
The reflection symmetry operators can take four possible forms: (A)$M_k=\pm\tau_0\otimes i\sigma_k$,  (B)$M_k=\pm\tau_z\otimes i\sigma_k$, (C)$M_k=\pm\tau_x\otimes i\sigma_k$and (D)$M_k=\pm i\tau_y\otimes i\sigma_k$($k=x,z$, see Appendix), among which (C, D) represent glide mirror symmetry.

As is shown in FIG. \ref{mirror}., physically there are two classes of mirrors, the mirror parallel to rotation axis and the mirror perpendicular to the rotation axis, respectively correspond to $k=x$ and $k=z$ above.

We can now give the physical interpretation of these mirror operators below, where $A/B$ represent two inequivalent atom sites (see FIG. \ref{mirror}), $|P_\pm\rangle = |P_x\rangle\pm i|P_y\rangle$, and $\uparrow,\downarrow$ represent spin:
\begin{enumerate}
\item We set the four basis as $(|P_{A+\uparrow} \rangle, |P_{A-\uparrow}\rangle , |P_{A-\downarrow}\rangle, |P_{A+\downarrow}\rangle)$. 
Then the $yz$-plane mirror operation interchange $P_+,P_-$ and $\uparrow,\downarrow$, i.e. the mirror operation $M_x:|P_{A+\downarrow}\rangle\rightarrow-i|P_{A-\uparrow}\rangle,|P_{A+\uparrow}\rangle\rightarrow-i|P_{A-\downarrow}\rangle, |P_{A-\uparrow}\rangle\rightarrow-i|P_{A+\downarrow}\rangle,|P_{A-\downarrow}\rangle\rightarrow-i|P_{A+\uparrow}\rangle$. The matrix representation of mirror operator is $M_x=-\tau_0\otimes i\sigma_x$. 

\item We set the four basis as $(|P_{A+\downarrow}\rangle, |P_{B+\downarrow}\rangle,-|P_{A-\uparrow}\rangle, -|P_{B-\uparrow}\rangle)$. A glide plane operator can have the following transformation $M_x:|P_{A+\downarrow}\rangle\rightarrow-i|P_{B-\uparrow}\rangle$, $|P_{B+\downarrow}\rangle\rightarrow-i|P_{A-\uparrow}\rangle$, $|P_{A-\uparrow}\rangle\rightarrow-i|P_{B+\downarrow}\rangle$,$|P_{B-\uparrow}\rangle\rightarrow-i|P_{A+\downarrow}\rangle$. Therefore the matrix representation of mirror operator is $M_x=\pm\tau_x\otimes i\sigma_x$.  

In general basis for this mirror operator can be constructed as following: There are two inequivalent sites(A and B) and the distance between them is $\vec t$. This transformation can be provided by defining: 
$|P_{A\pm\uparrow(\downarrow)}\rangle=e^{\pm i\vec r\cdot \vec K}u_{A\pm\uparrow(\downarrow)}, |P_{B\pm\uparrow(\downarrow)}\rangle=e^{\pm i M\vec r\cdot \vec K}u_{B\pm\uparrow(\downarrow)}$, where $\vec K$ denote the point in Brillouin zone, $u_{A\pm}(M\vec r+\vec t)e^{\pm i\vec t\cdot\vec K}=u_{B\pm}(\vec r)$ and $u_A(\vec r+M\vec t+\vec t)=u_A(\vec r)$. $M_z: \vec r\rightarrow M\vec r+\vec t$, where $M$ is a symmorphic mirror operation acting on $\vec k$ space and $\vec K\cdot(M\vec t+\vec t)=2\pi n$. For $\mathrm{BiO_2}$ of space group \#227, the Dirac point appears in the X point of its Brillouin zone with reflection symmetry belong to this case.
 
\item  We set the four basis as $(|P_{A+\uparrow} \rangle,|P_{A+\downarrow}\rangle , |P_{A-\downarrow}\rangle,- |P_{A-\uparrow}\rangle)$ $yz$-plane mirror operation $M_x:|P_{A+\downarrow}\rangle\rightarrow-i|P_{A-\uparrow}\rangle$, $|P_{A+\uparrow}\rangle\rightarrow-i|P_{A-\downarrow}\rangle$, $|P_{A-\uparrow}\rangle\rightarrow-i|P_{A+\downarrow}\rangle$,$|P_{A-\downarrow}\rangle\rightarrow-i|P_{A+\uparrow}\rangle$ has a  matrix form $M_x=-\tau_z\otimes i\sigma_x$. 

\item We set the four basis as $(|P_{A+\downarrow}\rangle,|P_{B+\downarrow}\rangle,-|P_{A-\uparrow}\rangle,-|P_{B-\uparrow}\rangle)$. The mirror operation on $xy$-plane writes $M_z:|P_{A+\downarrow}\rangle\rightarrow i|P_{B+\downarrow}\rangle,|P_{B+\downarrow}\rangle\rightarrow-i|P_{A+\downarrow}\rangle,|P_{A-\uparrow}\rangle\rightarrow-i|P_{B-\uparrow}\rangle,|P_{B-\uparrow}\rangle\rightarrow i|P_{A-\uparrow}\rangle$, with a matrix representation $M_z=i\tau_y\otimes i\sigma_z$. There are two inequivalent sites(A and B) and the distance between them is $\vec t$. This transformation can be provided by defining: 
$|P_{A\pm\uparrow(\downarrow)}\rangle=e^{\pm i\vec r\cdot \vec K}u_{A\pm\uparrow(\downarrow)}, |P_{B\pm\uparrow(\downarrow)}\rangle=e^{\pm i M\vec r\cdot \vec K}u_{B\pm\uparrow(\downarrow)}$, where $\vec K$ denote the point in Brillouin zone, $u_{A\pm}(M\vec r+\vec t)e^{\pm i\vec t\cdot\vec K}=u_{B\pm}(\vec r)$ and $u_A(\vec r+M\vec t+\vec t)=u_A(\vec r)$. $M_z: \vec r\rightarrow M\vec r+\vec t$, where $M$ is a symmorphic mirror operation acting on $\vec k$ space and $\vec K\cdot(M\vec t+\vec t)=(2n+1)\pi$.

\end{enumerate}

\subsection{The constraint on Hamiltonian by reflection symmetry and rotational symmetry}\label{CHRS}

Now we show how the Hamiltonian in $k_z$ axis is constrained by reflection symmetry and rotational symmetry:

Mirror plane parallel to $k_z$ axis: First we set the mirror plane as $yz$ plane and the mirror operator as $M_x$. The combination of $M_x$ and n-fold rotation operation generate n-1 more mirror planes and therefore all these mirror operators can be denoted as $M_k=C_n^kM_x(k=0,1,...,n-1)$. These mirror planes cross on $k_z$ axis and confine the Hamiltonian in $k_z$ axis. For each $M_k$, there's commutation relation $M_kH(k_z)M_k^{-1}=H(k_z)$. For the basis satisfying constraint relations (\ref{eqcrz}), when $M_x=\tau_0/\tau_z\otimes i\sigma_x$, the Hamiltonian becomes $H(k_z)=diag[a_0+a_{30},a_0-a_{30},a_0+a_{30},a_0-a_{30}]$; when $M_x=\tau_x/i\tau_y\otimes i\sigma_x$, the Hamiltonian becomes $H(k_z)=diag[a_0+a_{33},a_0-a_{33},a_0+a_{33},a_0-a_{33}]$. It is notable that at TRIM all $a$ terms with odd parity eliminated due to the relation $[H,T]=0$ and only $a_{00},a_{10},a_{30},a_{21},a_{22},a_{23}$ survive. Therefore, the constraints on $(p,q,r,s)$ are lessen in TBC.
All the results of the physical properties of Dirac semimetal are shown in TABLE \ref{tab:m}.

Mirror plane perpendicular to $k_z$ axis: First we set the mirror plane as $xy$ plane and the mirror operator as $M_z$. Along the $k_z$ axis, the reflection symmetry is $M_zH(k_z)M_z^{-1}=H(-k_z)$. Combined with $C_2, C_4, C_6$ rotational symmetry, reflection symmetry can create inversion symmetry $P=M_z\hat{C}_2$. The rotation operator for orbital space is $C_n=i\exp(i\pi \frac{1+p+q}{n})\exp(i\pi\frac{p-q}{n}\tau_z)=i\exp(i\pi \frac{1+p+q}{n})[\cos\frac{\pi}{n}(p-q)+i\tau_z\sin\frac{\pi}{n}(p-q)]$. Two-fold rotation is $\hat{C}_2=\cos\theta\tau_0+\sin\theta\tau_z$ where $\theta=\frac{\pi}{2}(p-q)$. If $p-q=0,2,4$, then $\hat{C}_2=\pm\tau_0$. These conditions can be satisfied when $p=q$ for $C_2$ symmetry, when $p=q$ or $p=q+2$ for $C_4$ symmetry, when $p=q$, $p=q+2$ or $p=q+4$ for $C_6$ symmetry. If $p-q=1,3,5$, then $\hat{C}_2=\pm\tau_z$. These conditions can be satisfied when $p=1,q=0$ for $C_2$ symmetry, when $p=q+1$ or $p=q+3$ for $C_4$ symmetry, when $p=q+1$, $p=q+3$ or $p=q+5$ for $C_6$ symmetry. 
The symmetry constraints with TRS along $k_z$ axis are $M_zH(k_z)M_z^{-1}=T H(k_z)T^{-1}$ and $PH(k_z)P^{-1}=T H(k_z)T^{-1}$. 
After $[H,T]=0$ at TRIM and constraint relations(\ref{eqcrz}) are applied, all situations protecting Dirac points are shown in TABLE \ref{tab:m2}, holding the same result with Yang and Nagaosa\cite{yn} when $P=\pm\tau_0,\pm\tau_z,\pm\tau_x$. 
Note that there are additional conditions for inversion symmetry $P\Gamma_\mu P^{-1}=\pm \Gamma_\mu(\mu=03,30,33)$\cite{yn}, the case for mixed mirror operator is thus eliminated. When $P=\pm i\tau_y$ the system can also generate Dirac points. For $C_3$ symmetry, the system does not have inversion symmetry. We only have symmetry constraints $m_zH(k_z)m_z^{-1}=T H(k_z)T^{-1}$ and $[H,T]=0$ at TRIM and the result is shown in TABLE \ref{tab:d3h} .

\subsection{$k\cdot p$ model}

In this section, we use the four band model to describe the effective Hamiltonian of material with point group $C_{4v}$ such as $\mathrm{Cd_3As_2}$. We write the $4\times4$ effective Hamiltonian generally as:
\begin{equation} \label{Hg}
H_{\text{eff}}=\sum_{i,j=0}^3{d_{ij}(k)\Gamma_{ij}}
\end{equation}
where $\Gamma_{ij}=\tau_i\sigma_j$. The basis of effective Hamiltonian are four angular momentum eigenstates $|+\frac{1}{2}\rangle,|+\frac{3}{2}\rangle,|-\frac{1}{2}\rangle$ and $|-\frac{3}{2}\rangle$.

Then we investigate the representations of $\Gamma$ matrices and k polynomials with three symmetric operations: the four-fold rotation along Z axis $\hat{C_4}$, the vertical reflection $\hat{m_v}$ and the dihedral reflection $\hat{m_d}=\hat{C_4}\hat{m_v}$.

The matrix operation of symmetric operators are: (1)$U(\hat{C_4})=R_{\frac{1}{2}}(\hat{C_4})\oplus R_{\frac{3}{2}}(\hat{C_4})$, (2)$U(\hat{m_v})=R_{\frac{1}{2}}(\hat{m_v})\oplus R_{\frac{3}{2}}(\hat{m_v})$, and (3)$U(\hat{m_d})=U(\hat{m_v})U(\hat{C_4})$, where $R_{j}(\hat{C_4})=\exp(i\frac{\pi}{2}j\sigma_z)$,$R_{j}(\hat{m_v})=\exp(i\frac{\pi}{2}j\sigma_x)$.

The operators act on the k polynomials $d_{ij}(k)$ as : (1)$\hat{C_4}: k_x\rightarrow -k_y,k_y\rightarrow k_x,k_z\rightarrow k_z$, (2)$\hat{m_v}: k_x\rightarrow -k_x,k_y\rightarrow k_y,k_z\rightarrow k_z$, (3)$\hat{m_d}: k_x\rightarrow k_y,k_y\rightarrow k_x,k_z\rightarrow k_z$.

\begin{table} 
\centering
\caption{The representations of $\Gamma$ matrices and k polynomials}
\label{trep}
	\begin{tabular}{l|l|l|l}
	\hline\hline
	\bf{Reps}	  & $\mathbf{\Gamma}$\bf{ matrices}		&$\mathbf{d(k)}$	& $\mathbf{T}$\\
	\hline
	$\tilde\Gamma_1$    &$\Gamma_{00},\Gamma_{30}$    & $1,k_x^2+k_y^2,k_z^2$ &+  	\\
	$\tilde\Gamma_1$    &-							& $k_z$				 &$-$	\\
	$\tilde\Gamma_2$    &$\Gamma_{03},\Gamma_{33}$    &  -           		 &$-$	\\
	$\tilde\Gamma_3$    &$\Gamma_{22}$         		& $k_x^2-k_y^2 $        &+		\\
	$\tilde\Gamma_3$    &$\Gamma_{12}$		  		& -					 &$-$	\\
	$\tilde\Gamma_4$    &$\Gamma_{21}$         		& $ k_x k_y   $         &+  	\\
	$\tilde\Gamma_4$    &$\Gamma_{11}$         		& -			          &$-$  	\\
	$\tilde\Gamma_5$    &$(\Gamma_{10},\Gamma_{23})$  &$(k_xk_z,k_yk_z)$		 &+ 		\\
	$\tilde\Gamma_5$    &$(\Gamma_{32},\Gamma_{01}),(\Gamma_{02},\Gamma_{31}),(\Gamma_{20},-\Gamma_{13})$        &$(k_x,k_y)$	&$-$ \\
	\hline
	\end{tabular}
\end{table}
	
The representations of $\Gamma$ matrices and k polynomials $d(k)$ are shown in TABLE \ref{trep}. By assembling the $\Gamma$ matrices and $d(k)$ with the same representation and time-reversal eigenvalue we obtain our effective Hamiltonian:
\begin{equation}\label{He}
\begin{split}
&H_{\text{eff}}(\vec k)=\epsilon(\vec k)+\\
&\left(\begin{array}{cccc}
A(\vec k) & -D(\vec k)k_+ & B_+k_- & -C(\vec k)\\ 
-D^*(\vec k)k_- & -A(\vec k) & C(\vec k) & B_-k_+\\
B_+k_+ & C^*(\vec k) & A(\vec k) & D(\vec k)k_-\\
-C^*(\vec k) & B_-k_- & D^*(\vec k)k_+ & -A(\vec k)
\end{array}\right)
\end{split}
\end{equation}
where $\epsilon(\vec k)=E_0+E_1k_+k_-+E_2k_z^2, A(\vec k)=A_0+A_1k_+k_-+\sqrt{A_2k_z^2+A_{20}^2}, B_\pm=\pm B_0+B_1, C(\vec k)=\frac{C_0}{2}(k_+^2+k_-^2)+\frac{i}{4}C_1(k_+^2-k_-^2), D(\vec k)=D_0+iD_1k_z$, and $k_\pm=k_x\pm ik_y$.


The Hamiltonian (\ref{He}) shows the possibility of coexistence of 3D Dirac points and Weyl line nodes. By choosing proper parameters shown in TABLE \ref{tpara}  (Only $B_0$ is different from $\mathrm{Cd_3As_2}$) we can verify our statement through the band structure FIG. \ref{fig:phase}.(a). FIG. \ref{fig:phase}.(a) is accomplished with the substitutions: $k_i\rightarrow \frac{1}{L_i}\sin(k_iL_i),k_i^2\rightarrow\frac{2}{L_i^2}[1-\cos(k_iL_i)]$ for a periodic lattice, where $L_x=L_y=a=12.67 \AA$ and $L_z=c=25.48\AA$\cite{ex1}.

\begin{table} 
\centering
\caption{Parameters for the 4$\times$4 effective Hamiltonian. Most parameters can be referred to from Jeon, Sangjun, et al\cite{ex2}.}
\label{tpara}
	\begin{tabular}{cccc}
	\hline\hline
	$E_0(eV)$		&	-0.219	&	$B_0(eV\AA)$		&	5	\\
	$E_1(eV\AA^2)$	&	-30		&	$B_1(eV\AA)$		&	0	\\
	$E_2(eV\AA^2)$	&	-16		&	$C_0(eV\AA^2)$		&	0	\\
	$A_0(eV)$		&	-0.060	&	$C_1(eV\AA^2)$		&	0	\\
	$A_1(eV\AA^2)$	&	18		&	$D_0(eV\AA)$		&	-2.75	\\
	$A_2(eV^2\AA^2)$&	96		&	$D_1(eV\AA^2)$		&	0	\\
	$A_{20}(eV^2)$	&	0.050		&	~&~\\
	\hline\hline
	\end{tabular}
\end{table}	

In order to solve Weyl line nodes, it is convenient to neglect the symmetric terms $\epsilon(\vec k)$ and some parameters $B_1,C_0,C_1,D_1$ as is shown in TABLE \ref{tpara}. 
Therefore, explicit calculation gives the equation of Weyl line nodes:
\begin{equation}
(B_0^2-D_0^2)k_x^2=\left[A_0+A_1k_x^2+\sqrt{A_2k_z^2+A_{20}^2}\right]^2
\end{equation}
This is schematically shown in FIG. \ref{fig:phase}.(c).

In FIG. \ref{fig:phase}, the $B_1,C_0,C_1,D_1$ terms are neglected. $B_0$ and $D_0$ are two main parameters determining whether there is a gap or Weyl line nodes in BZ. The phase diagram of $B_0$ and $D_0$ is shown in FIG. \ref{fig:phase}.(b). The critical line where Weyl line nodes appear and disappear is close to the line $D_0=B_0$, but $B_0$ is actually slightly larger than $D_0$, which can be shown more clearly with a smaller $A_0$ [see FIG. \ref{fig:phase}.(b) the dark blue area]. On the critical line, the two bands pull apart and the Weyl line nodes eventually annihilate accompanied by opening a gap.

\end{document}